\RequirePackage{lineno}
\documentclass[aps,prl,superscriptaddress,showpacs,twocolumn]{revtex4-1}
\usepackage{etoolbox}
\usepackage{graphicx}
\usepackage{floatflt}
\usepackage{float}
\usepackage{tabularx}
\usepackage{comment}
\usepackage{mathtools}

\usepackage{amssymb}

\usepackage[utf8]{inputenc}
\usepackage[T1]{fontenc}
\begin{document}



\title{Using the Neutron Fizeau Effect and Neutron Interferometry to Measure Energy-Dependent Contributions to the Neutron Optical Potential}


\author{V. Kurmangaliyeva}
\author{O. Agyl-Mussapar}
\author{S. Amangeldinova}
\author{B. Massak}

\affiliation{al-Farabi Kazakh National University, Almaty, Kazakhstan}
\author{D. Nassirova} 
\author{V. Zhumabekova}
\affiliation{Abai Kazakh National Pedagogical University, Almaty, Kazakhstan}
\author{W. M. Snow}
\affiliation{Indiana University/CEEM, 2401 Milo B. Sampson Lane, Bloomington, Indiana 47408, USA}

\date{\today}

\begin{abstract}
We propose a method to measure the energy dependence of the neutron optical potential $dV_{opt}/dE$ in the slow neutron energy regime. Our method makes essential use of a special property of the phase shift for a nonrelativistic neutron in moving matter, known as the neutron Fizeau effect. If a neutron traverses a medium which moves along the surfaces of its own parallel boundaries, the neutron only experiences a phase shift if the neutron optical potential of matter depends on the incident neutron energy. This feature of the neutron Fizeau effect can be combined with newly-developed forms of neutron interferometry to conduct sensitive measurements of $dV_{opt}/dE$. We describe some examples of scientific applications of this idea in the fields of neutron optics, subthreshold neutron-nucleus resonances, parity violation in low-energy p-wave neutron-nucleus resonances, and neutron scattering amplitudes of the nuclei of rare earth elements.      


\end{abstract}



\maketitle

\section {Introduction}

Neutron optics~\cite{Sea82, Searsbook} describes the neutron interaction with a medium in the forward scattering limit $\vec{q} \to 0$, where $\vec{q}$ is the momentum transfer, and is based on the coherent state formed by the incident wave and the forward scattered wave in a scattering medium~\cite{Gol64}. In the one-body Schrodinger equation for the neutron motion

\begin{equation}
     \label{eq:coherent}
     \left[\frac{-\hbar^{2}}{ 2m}\Delta +V(r)\right]\psi(r)=E\psi(r)
 \end{equation}
 $\psi(r)$ is the coherent wave and $V(r)$ is the optical potential of the medium. In the kinematic approximation which neglects the effects of dispersion and multiple scattering in the medium, the neutron optical potential in a medium of constant density becomes

\begin{equation}
     \label{eq:opticalpotential}
     V_\mathrm{opt}~=~(2\pi \hbar^2/m)\sum_{l} N_{l} b_{coh, l} (0,E) 
     =V_{opt}(E),
\end{equation}

where $N_{l}$ is the number density of scatterers, $m$ is the neutron mass, and $b_{coh}(0,E)$ is the coherent scattering amplitude for a neutron of energy $E$ with momentum transfer $\vec{q}=0$. Recall that the coherent scattering amplitude $b_{coh}(0,E)=({{I+1} \over {2I+1}})b_{+} + ({{I} \over {2I+1}})b_{-}$ consists of the linear combination of scattering amplitudes $b_{+}$ and $b_{-}$ for the two angular momentum channels $J=I\pm 1/2$ in s-wave neutron-nucleus scattering that leave the state of the scattering medium unchanged, and is therefore the amplitude that appears in neutron optics. The neutron index of refraction is then

\begin{equation}
     \label{eq:oldindex}
     n^{2}=1-\frac{V_{opt}(E)}{E}
 \end{equation}

and if in addition $b_{coh}(0,E)=b_{coh}$ is a constant independent of neutron energy, then the neutron optical potential is energy-independent: $V_{opt}(E)=V_{opt}$. 

A nontrivial energy dependence for $V_{opt}(E)$ can come from two main sources. The relation between the neutron optical potential and the scattering amplitude described above is an approximation which neglects effects due to dispersion and multiple scattering in the medium. Both of these sources come from the effects of the medium and can be present even if $b_{coh}(0,E)$ has no energy dependence.   

$V_{opt}(E)$ can also depend on energy through the energy dependence of $b_{coh}(0,E)$. In the neutron-nucleus interaction, one encounters both potential scattering and resonance scattering. In the $kR<<1$ limit of slow neutrons where $k$ is the neutron wave vector and $R$ is the nuclear radius, the potential scattering gives a constant amplitude independent of neutron energy to high precision. Neutron-nucleus resonances also exist, with resonance energies located both above and below the neutron separation energy $S_{n}$ of the $A+1$ system formed as the neutron interacts with a nucleus with nucleon number $A$. A neutron with kinetic energy $E$ thereby creates an excited state in the $A+1$ system of energy $S_{n}+E$. Resonances in this $A+1$ system can therefore be present both for $E>0$ and also for $E<0$: we will refer to the latter resonances as negative energy resonances as is common in the neutron physics scientific literature. In heavy nuclei the density of energy levels is high enough that it is not uncommon to encounter such resonances, both positive and negative, with a resonance energy $E_{r}$ close to zero. $b_\mathrm{atom}(0,E)$ can then possess a nontrivial dependence on $E$.


In both of these cases, to perform sensitive measurements of the energy dependence of $V_{opt}(E)$ it would be especially useful to employ a measurement method which is insensitive to the neutron-nucleus potential scattering process, which gives a negligible contribution to the energy dependence of the neutron-nucleus scattering amplitude in the thermal neutron energy regime. The quantity one would ideally want to measure in this case is the derivative $dV_{opt}(E)/dE$. 

To realize this idea one needs a method to measure the coherent neutron scattering amplitude as a function of neutron energy. Neutron interferometric measurements of the relative phase of two neutron amplitudes can achieve the high sensitivity required for this goal. The ability of slow neutrons to penetrate macroscopic amounts of matter and to interact coherently with the medium allow the quantum amplitudes governing their motion to accumulate large phase shifts which can be sensed with interferometric measurements~\cite{Nico05b, Dubbers11, Pignol:2015}. Already neutron interferometry combined with measurements using other neutron optical methods have determined the coherent neutron scattering amplitudes $b_{coh}$ at thermal neutron energies at $10^{-3}$ precision for several nuclei~\cite{Koester1991}. For cases where precision neutron interferometric measurements of coherent scattering amplitudes have been conducted using independent techniques, the internal consistency of the results to date is high~\cite{Snow2020}. 

Recent developments in neutron interferometry now enable one to realize precision neutron coherent scattering amplitude measurements as a function of neutron energy $E$ for a range of slow neutron energies near $E=0$. In this paper we propose to employ these new neutron interferometric methods in combination with an interesting property of the neutron Fizeau effect in moving matter to isolate the neutron energy-dependent part of the neutron optical potential, $dV_{opt}(E)/dE$. Previous work in neutron optics theory and experiment has shown that the phase shift of a neutron in a moving medium, in the particular case where the medium boundaries are flat and parallel and where the motion of the medium is parallel to the boundaries, is directly proportional to $dV_{opt}/dE$. This feature of the neutron Fizeau effect was verified in the past in a pioneering measurement of a low energy neutron-nucleus resonance in $^{149}$Sm using a perfect crystal neutron interferometer.




The rest of this paper is organized as follows. We first review the neutron interferometry method used for precision measurement of the neutron-nucleus coherent scattering amplitude $b_{coh}$. We then review the physics behind the neutron Fizeau effect and the results of the only previous measurement to our knowledge which used this technique to determine neutron-nucleus resonance parameters. We then review the theory of neutron-nucleus scattering in the isolated resonance regime, including the contribution from both positive and negative neutron-nucleus resonances. Here we must take care to properly treat certain special features of the resonant contribution to the scattering amplitude close to threshold, which is the limit of interest to our work. Fortunately we can draw on previous work in nonrelativistic scattering theory for this purpose. This is followed by discussions of various examples of scientific issues which such a measurement method for $dV_{opt}(E)/dE$ can address. Finally we discuss various possible sources of systematic error associated with this measurement method. 

\section{Neutron Interferometric Methods for Neutron Optical Potential Measurement}

Neutron interferometry is described in great detail in a recent book~\cite{RauchWerner}. Historically the first sensitive neutron interferometry measurements were conducted using perfect crystals to split and recombine the neutron beam, and we use this example to explain the technique. Perfect crystal neutron interferometric measurements of scattering amplitudes employ a Mach-Zehnder interferometer in which the neutron amplitude $\psi e^{-i\Phi}$ is coherently split into two paths. The measured phase shift is dominated by the real part of the neutron optical potential $V(x)$ and can be expressed as~\cite{RauchWerner} 


\begin{equation}
    \label{eq:Normalphaseshift}
   \Phi={m \over k \hbar^{2}}\int{V(x)_\mathrm{opt}dx},
\end{equation}

\noindent where $m$ is the neutron mass, $k$ is the neutron wave number, and $V(x)_\mathrm{opt}$ is the neutron optical potential at location $x$. A neutron moving through the medium acquires a phase shift $\Delta \Phi$ of the form 



\begin{equation}
    \label{eq:phase}
    \Delta \Phi = (n_{r} - 1) k D = - \sum_{l} \lambda_{n} N_{l} b_{l} D,
\end{equation}

where $D$ is the thickness of the sample medium along the direction of neutron propagation and $\lambda_{n}$ is the neutron wavelength. A perfect crystal neutron interferometer consists of three crystal blades on a common crystal base. The first blade serves to spatially separate the neutron's wave function $\psi e^{-i\Phi}$ into two coherent paths (A and B). In order for the two paths interfere a central crystal blade directs the paths back together onto the third blade, where the paths interfere. Neutrons exit the interferometer along either one of two paths labeled traditionally as  `O' and `H'  and are detected using highly efficient $^3$He-filled proportional counters. Differences in phase $\Delta\Phi$ between the paths A and B modulate the intensities recorded by the detectors as
\begin{eqnarray}
\label{eqn1}
I_O = A_O + B\cos[\xi(\delta)+\Delta\Phi] \label{IO} \\
\label{eqn2}
I_H = A_H - B\cos[\xi(\delta)+\Delta\Phi] \label{IH}
\end{eqnarray}
\noindent 

In order to determine $\Delta\Phi$ and the other fit parameters ($ A_{O,H}$ and $B$) one varies the cosine term in ($\xi(\delta)$) by the adding a `phase flag' inside the interferometer. By rotating the phase flag by an angle $\delta$ an adjustable phase shift of $\xi(\delta)$ is introduced between paths A and B to form an interferogram which can be used to measure the phase shift. 

The dynamical diffraction in perfect crystals used in perfect crystal neutron interferometer beamsplitters greatly restricts the phase space acceptance of the device to narrow slices of neutron energy close to those set by the crystal diffraction condition. Three new neutron interferometry techniques developed recently enable one to measure the neutron phase shift both with higher statistical accuracy and also over a broader range of slow neutron energies, thereby enabling sensitive measurements of $dV_{opt}(E)/dE$ over a wide range of neutron energies. One new neutron interferometer device~\cite{Fujie2024} employs neutron beamsplitting etalons to split and recombine the neutron paths. Since these devices are based on mirror reflection, they can be used over a broad range of slow neutron energies.  Other methods take advantage of recent advances in neutron phase gratings to develop Moire neutron interferometers, which like the etalon device mentioned above can accept a broad range of neutron energies in the slow neutron regime~\cite{Sarenac2018}, and neutron interferometers based on the Talbot-Laue effect~\cite{Clauser1994, Pfeiffer2006, Momose2020}. Our idea could be realized in various ways with any of these interferometer types.

\section{Neutron Fizeau Effect Review}

Our idea exploits a somewhat surprising fact about the neutron optics version of the famous Fizeau effect, the extra phase shift of light in a moving medium first demonstrated in the 19th century. When the neutron optics version of the Fizeau effect was analyzed theoretically more than a century later~\cite{Horne83} it was quickly discovered that the nonrelativistic energy-momentum dispersion relation for slow neutrons leads to different motion-induced phase shifts than for light. In contrast to the Fizeau effect for light, for example, there is no predicted phase shift for a slow neutron as it passes through a medium with flat boundaries, if (1) the motion of the medium is parallel to the boundaries, and (2) the neutron optical potential of the medium does not depend on neutron energy. 

Below we briefly review the theory and experiments on the neutron Fizeau effect to set the context for our idea. To our knowledge almost all of the work performed to date on this subject is well summarized in a book on neutron interferometry~\cite{RauchWerner} whose presentation based on~\cite{Horne83} we largely follow. Consider a neutron of wave vector $\vec{k}$ incident on a plate with parallel boundaries of thickness $L$ composed of a uniform medium with neutron index of refraction $n(\vec{k})$. If the plate is at rest, the phase shift of the neutron induced by the medium is

\begin{equation}
\label{eqn3}
\phi(\vec{k})= (K_{x}-k_{x})L=(\sqrt{n^{2}(\vec{k})k^{2}-k_{y}^{2}}-k_{x})L
\end{equation}
\noindent   

where $k_{x}$ and $k_{y}$ are the components of the neutron wave vector normal to and along the surface of the plate, respectively, and $\vec{K}=n(\vec{k})\vec{k}$ is the wave vector inside the medium. If the plate is in motion in the lab frame with velocity $\vec{w}$, the phase shift  will be $\phi(\vec{k'})$  in the rest frame of the plate, where $\vec{k'}$ is the incident wave vector as seen in the moving frame. The relativistic invariance of the phase implies that, for uniform motion of the plate, the phase shift induced by the motion (the Fizeau effect) is 

\begin{equation}
\label{eqn3}
\Delta \phi=\phi(\vec{k'})-\phi(\vec{k})
\end{equation}
\noindent  

and depends both on the functional form of the neutron index of refraction $n(\vec{k})$ and the relation between $\vec{k'}$ and $\vec{k}$ from the Lorentz transformations. Conservation of energy for nonrelativistic neutron motion in the optical potential $V_{opt}(|\vec{k}|)$ of the medium gives 

\begin{equation}
\label{eqn3}
n(\vec{k})=K_{x}/k_{x}=\sqrt{1-{2mV_{opt}(|\vec{k}|) \over \hbar^{2}k^{2}}}
\end{equation}
\noindent  

In the nonrelativistic limit the Galilean transformation law $\vec{k'}=\vec{k}-m\vec{w}/\hbar$ holds, and one can express the Fizeau phase shift as

\begin{equation}
\label{eqn3}
\Delta \phi=-((n^{2}-1)+k^{2}{\partial n^{2} \over \partial k^{2}})mwL\tan{\theta}/\hbar
\end{equation}
\noindent 

to first order in $w=|\vec{w}|$, where $\theta$ is the angle between $\vec{k}$ and the normal to the plate.  If the coherent neutron-nucleus scattering amplitude $b$ is dominated by s-wave scattering so that the neutron optical potential takes the usual form $V_{opt}={2\pi \hbar^{2}b \over m}$, then $dV/dE=0$ and there is no Fizeau phase shift for slow neutrons in this case. In the neutron case the kinematic and dispersive contributions to the phase shift cancel exactly in this limit: for light the kinematic term is dominant. The absence of a neutron Fizeau effect in this situation can be thought of in the following way: for the component of the neutron momentum normal to the direction of motion of the medium, the effective potential seen by the neutron is the same as when the medium is at rest as long as the neutron optical potential is energy-independent. 

These theoretical predictions were verified long ago in a series of measurements involving  phase shifts of a rotating quartz rod behind a two-slit neutron interferometer~\cite{Klein81}, a rotating aluminum propellor entraining both arms of a perfect crystal neutron interferometer~\cite{Bonse86}, and a spinning disk of quartz in a perfect crystal interferometer~\cite{Arif85}, which was used to verify the null neutron Fizeau effect.  In this latter experiment a quartz disk of neutron optical potential $V$ and thickness $L$ with faces parallel to the interferometer blades that overlapped both subbeams was spun so that the quartz moved with a speed $w$ with respect to the neutron subbeams in the interferometer, which make angles $\pm \theta$ with respect to the normal to the disk surface. In this geometry one can evaluate the Fizeau phase shift $\Delta \phi$ in the perfect crystal neutron interferometer to be

\begin{equation}
\label{eqn4}
\Delta \phi=({-2m_{n}w \over \hbar}){dV \over dE} L\tan{\theta}
\end{equation}
\noindent 

where $m$ is the neutron mass, $w$ is the speed of the motion of the medium, and $\theta$ is the angle between the incident neutron momentum and the normal to the surface of the moving medium. The diffracted subbeams in the first blade of the perfect crystal neutron interferometer generated a large enough angle $\theta$ relative to the normal to the disk to make $\Delta \phi$ linear in $w$ and to place the same spinning disk in both subbeams. In the case of a flat quartz disk moving parallel to its boundaries, no Fizeau effect was seen as expected, since the neutron-nucleus interactions with the silicon and oxygen nuclei in quartz possess no low-energy neutron-nucleus resonances and the potential scattering at low energy is dominated by s-wave scattering which leads to an energy-independent neutron optical potential.

In the pioneering measurement of Arif et al.~\cite{Arif89}, a clear nonzero $dV/dE$ from a low energy neutron-nucleus resonance in $^{149}$Sm was measured using a perfect crystal neutron interferometer. The measurement setup was very similar to that of the quartz disk experiment~\cite{Arif85}, but in this case the measurement employed a spinning disk of $^{149}$Sm whose angular velocity was changed both in sign and in magnitude. The axis of the disk was in the plane of the incident neutron beam, and the spinning wheel extended over both interferometer subbeams. The nonzero phase shifts agreed with the computed value of $dV/dE$ using the known parameters of the resonance. 

The measurement of Arif et al. was performed at a single neutron energy of $95.8$ meV, close in energy to the $97.3$ meV resonance in $^{149}$Sm, 13\% abundant in natural samarium. Even in a very thin $33$ micron foil of natural samarium, one could still measure a $0.05$ radian phase shift with a disk rotation frequency of $300$ Hz. The ability of the various types of broadband neutron interferometry mentioned above to measure phase shifts over a much broader range of neutron energies would enable one to map out $dV/dE$. 

In this paper we propose to leave the discussion of the different types of implementations of our idea for different neutron interferometry methods to a later work. In all cases one needs to engineer a version of the conditions met in the perfect crystal interferometer measurement mentioned above: a moving uniform density mass in the form of a plate or disk that moves with a velocity parallel to its boundaries, either by rotation or oscillation, which generates a differential phase shift in the relevant pair of interferometer paths inside the device.  For the rest of this paper we restrict ourselves to three items: (1) a description of a proposed analysis model that takes into account both neutron potential and resonance scattering, including both subthreshold resonances and the special modifications to the neutron scattering amplitude for resonances close to $E=0$, (2) examples of scientific questions in different physics subfields which a device of this type can address, and (3) some potential sources of systematic error.

\section{Inclusion of Subthreshold and Near-Thershold Resonances into Neutron Optical Theory}

In the $q \to 0$ limit of coherent forward scattering relevant for neutron optics, $b_{coh}(0,E)$ possesses two qualitatively different contributions in the case of heavy nuclei: $b_{coh}(0,E)=b_{pot}+b_{res}(E)$. $b_{pot}$ is the potential scattering from so-called direct neutron-nucleus reactions. In the limit $kR<<1$ where $k$ is the incident neutron wave vector and $R$ is the radius of the nucleus, $b_{pot}$ is dominated by s-wave scattering and is therefore a constant that depends on the nucleus but is independent of neutron energy to an excellent approximation in our energy range of interest. $b_{res}(E)$ is the term due to resonances in the excited $A+1$ compound nuclear system. 

In the limit $kR<<1$ of interest for this work the widths of these resonances are typically narrow compared to their separation. Using the partial wave expansion in nonrelativistic scattering theory~\cite{Landau1977}, one gets the usual Breit-Wigner form for elastic resonance scattering

\begin{equation}
     \label{eq:BreitWigner}
 b_{res}=2ikf_{l}={{\Gamma/2}\over{E-E_{r}+i\Gamma/2}} 
 \end{equation} 

with a resonance energy $E_{r}$ and width $\Gamma$. In our case with inelastic channels from gamma emission the expression for the resonant part $b_{res}$ of the total scattering amplitude $b=b_{pot}+b_{res}$ becomes~\cite{Lynn1968, Landau1977}

\begin{equation}
    \label{eq:opticalpotential}
    b_{res}=\sum_{j}{g_{\pm, j} \over 2k^{'}_{j}}{\Gamma_{n,j} \over [(E^{'}-E_{j})+i\Gamma_{j}/2]}
\end{equation}

\noindent where $\Gamma_{n,j}$ and $\Gamma_{j}$ are the neutron width and total width of the resonance at energy $E_{j}$ and $k^{'}=\mu k/m$ is the wave vector in the n-A center of mass system of reduced mass $\mu$,  $E^{'}$ is the associated energy in the COM frame, and $g_{+, j}=(I+1)/(2I+1)$ and $g_{-,j}=I/(2I+1)$ are the statistical weight factors for a resonance at energy $E_{j}$ in the total angular momentum channel $J=I \pm 1/2$. $b_{res}$ is purely imaginary on resonance, as can easily be seen explicitly by writing out the real and imaginary parts of $b_{res}$

\begin{equation}
    \label{eq:opticalpotential}
    b_{res, real}=\sum_{j}{g_{\pm, j} \over 2k^{'}_{j}}{\Gamma_{n,j}(E^{'}-E_{j}) \over [(E^{'}-E_{j})^{2}+\Gamma_{j}^{2}/4]}
\end{equation}

\begin{equation}
    \label{eq:opticalpotential}
b_{res, im}=-\sum_{j}{g_{\pm, j} \over 4k^{'}_{j}}{\Gamma_{n,j}\Gamma_{j} \over [(E^{'}-E_{j})^{2}+\Gamma_{j}^{2}/4]}
\end{equation}

The complete expression for the low energy scattering amplitude $b$ then becomes, for positive energy resonances,

\begin{equation}
    \label{eq:opticalpotential}
    b=b_{pot}+b_{res}= b_{pot}+\sum_{j}{g_{\pm, j} \over 2k^{'}_{j}}{\Gamma_{n,j} \over [(E^{'}-E_{j})+i\Gamma_{j}/2]}
\end{equation}

and this is the usual expression that is used in fitting neutron-nucleus scattering amplitudes as a function of energy in the isolated resonance regime. 

However it is important to recognize that this expression does possesses an implicit assumption that the lowest resonance energy $E_{j}$ in the sum above is not too close to the point $E=0$. If $E_{j}$ is too close to zero, then the expression above varies rapidly as a function of $E$ near threshold. This contradicts the general result from scattering theory from a short-range potential, which says that the s-wave scattering that dominates the scattering amplitude in this limit should tend to a constant as $E \to 0$. As will be seen below, this situation is not merely of academic interest as it is known to happen in several cases for real nuclei. In addition, there is the separate question of how to treat the subthreshold resonances with $E_{j}<0$. 

Although the general treatment of this problem is involved, we can present useful formulae in the case that the resonance near $E=0$ has angular momentum $L=0$. In this case the correction to the equation above which gives a energy-independent scattering amplitude in the $E \to 0$ limit is known.




The standard cure is the Landau/Wigner threshold law~\cite{Landau1977}:
for an open $L=0$ s-wave channel the neutron partial width scales as
$\Gamma_{n}(E)\propto k\propto\sqrt{E}$, which restores a finite $E\!\to\!0$ limit
for the amplitude. This refinement is essential for our Fizeau-based
measurement of $dV_{opt}/dE$, since the phase shift vanishes unless
$V_{opt}$ depends on $E$. To accommodate this case we separate the near-threshold $s$-wave level $j\!=\!0$ from all other more distant levels:


\begin{equation}
  b(E)=b_{\mathrm{pot}}+b_{\mathrm{far}}(E)+b_0(E)\,
  \label{eq:split1}
\end{equation}

where $k'=\sqrt{2\mu E'}/\hbar$ refers to the $n$–$A$ channel, and $b_{\mathrm{far}}(E)$ is the usual expression

\begin{equation}
 b_{\mathrm{far}}(E)=\sum_{j\neq 0}\frac{g_{\pm,j}}{2k'_j}\,
  \frac{\Gamma_{n,j}}{(E'-E_j)+\mathrm{i}\,\Gamma_j/2}\,,
    \label{eq:split2}
\end{equation}


For the single $s$-wave level near $E=0$, we use the Wigner law
$\Gamma_{n,0}(E)=\beta_0\,k'$ and
$\Gamma_0(E)=\Gamma_{\gamma,0}+\Gamma_{n,0}(E)$ to obtain the corrected term
\begin{equation}
  b_{0}(E)=\frac{g_{\pm,0}}{2k'}\,
  \frac{\Gamma_{n,0}(E)}{(E'-E_0)+\mathrm{i}\,\Gamma_0(E)/2}\,.
  \tag{18a}
\end{equation}
Equivalently, one may absorb smooth factors into constants
$\mathcal C_0,\alpha_0>0$ and write
\begin{equation}
  b_{0}(E)=\frac{g_{\pm,0}}{2k'}\,
  \frac{\mathcal C_0}{\,E'-E_0+\mathrm{i}\,\alpha_0\sqrt{E'}\,}
  \label{eq:landau_den}
\end{equation}
which is the $(E-\varepsilon_0+\mathrm{i}\,\gamma\sqrt{E})^{-1}$ denominator discussed by Landau near a quasi-discrete level and guarantees a constant $s$-wave limit as $E\!\to\!0$. The two parameterizations are equivalent under a simple redefinition of $\{\mathcal C_0,\alpha_0\}$.


As it turns out this same construction also works as well for a subthreshold resonance near $E=0$. Analyticity at the branch point~\cite{Landau1977, Taylor2006} requires $\sqrt{E'}\!\to\!\mathrm{i}\sqrt{|E'|}$ and
$k'\!\to\!\mathrm{i}\kappa'$ with $\kappa'=\sqrt{2\mu|E'|}/\hbar$. The open neutron channel then closes so that the near-threshold term becomes purely real:
\begin{equation}
  b_{0}(E<0)=\frac{g_{\pm,0}\,\beta_0/2}{(E'-E_0)+\mathrm{i}\,\Gamma_{\gamma,0}/2}
  \in\mathbb{R}
  \tag{18b}
\end{equation}
or, in the Landau-style notation,
\begin{equation}
  \bigl[b_{0}(E)\bigr]_{E'<0}=
  \frac{g_{\pm,0}}{2k'}\,
  \frac{\mathcal C_0}{\,E'-E_0-\alpha_0\sqrt{|E'|}\,}\in\mathbb{R}.
\end{equation}

We therefore propose the form of equation~\ref{eq:split1} for analysis which properly treats the case where the resonance energies, positive or negative, are close to $E=0$. Additional negative energy resonances which are farther away from the $E=0$ threshold, as long as they also possess spacings which put them in the isolated resonance regime, can be modeled using the usual form for positive energy resonances and widths, but with $E_{j}$ negative. This approach is employed in global analyses of neutron-nucleus scattering data. More complicated behaviors are also possible in principle~\cite{Demkov1966}.   



Although the frequency,  amplitudes, and widths of $L=1$ neutron-nucleus resonances are all smaller than those of $L=0$ resonance due to the angular momentum barrier for penetration of the neutron wave function into the nucleus, nothing prevents a $L=1$ resonance from appearing very close to $E=0$. In fact there is a speculation in the scientific literature that precisely this situation may happen in an isotope of lead to cause a large parity-odd effect observed in neutron spin rotation~\cite{Heckel1982, Abov1989, Lobov2000, Andrzejewski2004, Oprea2014}. Resonances with $L=1$ and higher will possess a different energy dependence near threshold. 

In the next few sections we discuss various examples of scientific issues which the device and method we have described above would be able to address.  



\section{Subthreshold Neutron-Nucleus Resonance Parameters for Nuclear Data}

The properties of subthreshold neutron-nucleus resonances located just below the neutron separation energy of the $A+1$ nucleus formed in neutron capture on a $A$-nucleon system  are important to constrain for various applications in science and technology. The tail of the subthreshold resonance which extends above the neutron separation energy and into the thermal neutron energy range gives an added contribution to the neutron cross section on top of the potential scattering and the above-threshold resonance scattering contributions. Although extensive data exists for the energies and widths of the neutron-nucleus resonances above the neutron separation energy, the contributions from negative energy resonances below threshold are obviously inaccessible to direct measurement. Although $(d,p)$ reactions can access these same states experimentally, the precision required for the application of interest for neutron physics in the thermal neutron energy regime is not usually sufficient for nuclear data purposes. 

Extensive data on n-A resonances exists and is organized in several centralized nuclear data sources~\cite{Mughabghab2018a, Mughabghab2018b, NNDC, Shibata2011, BROND, Koning2006, CENDL}.  Nuclei which possess significant subthreshold effects can be isolated by inspecting the experimental data on the relevant n-A cross sections at low (meV) energies. Almost all of the data used in these analyses of subthreshold resonances comes from various types of neutron cross section data (mainly the total cross section and the (n, $\gamma$) cross section for non-fissile nuclei).

Existing methods to infer the resonance parameters of subthreshold neutron-nucleus resonances therefore mainly employ measurements of the energy dependence of neutron cross sections using a R-matrix analysis~\cite{Wigner1947, Lane1958}. The precision of the measurement of thermal neutron cross sections of the above-threshold resonances is usually high enough to be sure that the remaining observed variations of the cross section near $E=0$ must be coming from subthreshold resonances. In these cases it is common to add a subthreshold s-wave resonance with an adjustable resonance energy and width to fit the data~\cite{Bouland2007, Frohner2000, Aleksejevs1998}. The parameters of these subthreshold neutron-nucleus resonances are reported in nuclear data compilations and are updated periodically in the wake of improved neutron measurements of the above-threshold reactions. In view of the limitations on neutron reaction measurement precision and also the fact that there are typically many subthreshold resonances in most nuclei, the parameters one finds in the data tables for the properties of these negative energy resonances must be considered as effective parameters which encode information on the above-threshold tail of the full collection of negative energy resonances and not simply on the one closest to threshold. 

Typical precision of this data in the thermal neutron energy regime is at the 1\% level. It could be possible to improve our experimental knowledge of effects from subthreshold resonances if one could conduct measurements with higher precision as a function of neutron energy in the thermal neutron regime using the techniques we describe above.

\section{Subthreshold Neutron-Nucleus Resonance Parameters for Resonance Enhancement Near P-Wave Resonances}

Parity violation in p-wave resonances in heavy nuclei can be amplified as much as 6 orders of magnitude above the size expected by dimensional analysis~\cite{Sushkov1980, Sushkov1982, Flambaum1984, Flambaum1985, Alfimenkov1981, Alfimenkov1983, Shimizu1993, Alfimenkov1984, Mitchell1999, Mitchell2001}. The theory behind the greatly-amplified size of both parity-even and parity-odd effects measured in the neutron spin-dependent cross section and in the angular distribution of emitted gammas near certain low energy p-wave resonances in heavy nuclei at low energy has been known for decades~\cite{Flambaum1984, Flambaum1985}. Here we consider the case of parity violation in the total cross section. A nonrelativistic neutron under the influence of the strong and weak interactions can be described by a Hamiltonian of the form

\begin{equation}
H=p^{2}/2m +V(r)+{1 \over 2} \{ F(r), {\vec{\sigma} \cdot \vec{p}}  \} 
\end{equation}

and since the weak interaction violates parity, one can search for parity-odd effects in neutron-nucleus interactions. The amplification of parity-odd effects is largest on p-wave resonances in heavy nuclei, whose dense set of resonance levels can lead to large P-odd asymmetries in the total cross section from resonance-resonance mixing of size

\begin{equation}
A={{\sigma_{+}-\sigma_{-}} \over {\sigma_{+}+\sigma_{-}}}=2\Sigma_{s} 
{<s|V|p> \over (E_{s}-E_{p})} {\sqrt{\Gamma_{s} \over \Gamma_{p}}}
\end{equation}

where $A$ is the parity-odd asymmetry in the total cross section, $\sigma_{\pm}$ is the total cross section for $\pm$-helicity neutrons, $E_{s}$ and $E_{p}$ are the s-wave and p-wave resonance energies, $\Gamma_{s}$ and $\Gamma_{p}$ are the s-wave and p-wave resonance widths, and the sum is over all s-wave resonance with the same total angular momentum $J$ as that of the p-wave resonance. This asymmetry takes the maximum value shown above for an incident neutron energy $E=E_{p}$. 

In certain heavy nuclei where large P-odd effects have been observed (the $0.73$ eV p-wave resonance in $^{139}$La, the $0.88$ eV p-wave resonance in $^{81}$Br, the $1.33$ eV p-wave resonance in $^{117}$Sn, and others), these p-wave resonances are so close to neutron threshold and so far away from the positive s-wave resonances in these nuclei that the only-known quantitative explanation for the large size of the observed large P-odd effects relies on parity mixing of the p-wave resonance with a subthreshold s-wave resonance. In this case, the measurement approach outlined above can be useful for an improved quantitative understanding of the theory of the amplification of parity violation in heavy nuclei for these cases where mixing with a sub-threshold s-wave resonance dominates the asymmetry. 

This improved understanding is in turn also important for future experimental efforts to use these same systems to conduct sensitive searches for time reversal violation in polarized neutron transmission through polarized $^{139}$La, which is one of the cases where the amplification comes from interference with a subthreshold s-wave resonance~\cite{Bowman:2014fca, Endo2023, Nakabe2024}. The goal of this work is to search for a term in the neutron forward scattering amplitude of the form $\vec{\sigma}_n \cdot (\vec{k}_n \times \vec{I})$, where $\vec{\sigma}_{n}$ is the spin of the neutron, $\vec{k}_{n}$ is the neutron momentum, and $\vec{I}$ is the spin of the nucleus. This observable is both parity odd and time reversal odd. With the recent development of MW-class short pulsed spallation neutron sources and with advances in polarized $^{3}$He neutron spin filter technology for eV energy neutrons, the statistical accuracy of such a search for P-odd and T-odd interactions can have a sensitivity of about one order of magnitude beyond the present upper bound on the electric dipole moment of the neutron, and it is strongly suspected that such a search is likely to have a different sensitivity to various types of possible beyond-Standard-Model T-odd physics compared to electric dipole moment searches in nucleons and nuclei. A more accurate determination of the subthreshold resonance parameters in this system and in others using low-energy p-wave resonances with large parity violation would be valuable to help quantify and control certain forms of systematic error in the measurement. The work of Endo et al~\cite{Endo2023} presents the latest summary of the data on the resonances parameters close to $E=0$ in $^{139}$La. 


\section{Improved Neutron-Nucleus Scattering Amplitude Expressions for Neutron Scattering from Rare Earth Nuclei}

Polarized neutron scattering is one of the most powerful methods to investigate the internal structure and dynamics of magnetic materials. The understanding of magnetism involving rare earth elements is a major theme in materials science in this century in view of the extensive technological applications of these materials. Magnetic systems involving rare earth elements are also strong candidates to exhibit quantum many-body phenomena which go beyond the 20th-century Landau paradigm for the theoretical understanding of quantum many body systems. In these materials polarized neutrons interact both with the nucleus and also through the magnetic interaction of the neutron magnetic moment with the internal magnetization from the unpaired electrons. Since the size of these scattering amplitudes are comparable, one can get large constructive and destructive interference effects in certain elastic scattering processes. It is therefore important to understand the slow neutron-nucleus scattering amplitudes from these elements. 

By coincidence, it turns out that many of the isotopes of the nuclei among the rare earth elements also happen to possess neutron-nucleus resonances very close to threshold and in the slow neutron regime of interest to this work. Lynn~\cite{Lynn1989} and then Lynn and Seeger~\cite{Lynn1990} calculated the neutron energy dependence of the cross sections and the real part of the coherent scattering lengths in the isotopes of these elements where low energy resonances result in a complicated energy dependence. Both these works along with a recent update by Von Dreele~\cite{VonDreele2024} note that a negative energy resonance close to threshold is required to explain the data in the case of $^{151}$Eu.  

The precision with which these quantities could be inferred was limited by the data available at the time. A direct measurement of $dv_{opt}(E)/dE$ for these nuclei could determine the quantities of interest for neutron scattering from these elements with much higher precision. With the higher precision it could be important in some cases to employ the slightly-modified expression for the neutron scattering amplitude we discussed above.


\section{Local Field Effects in the Neutron Optical Potential}

Shortly after the neutron Fizeau experiments were conducted, Sears pointed out~\cite{Sears85} that the technique we have described above could be used to search for corrections to the kinematic theory of neutron optics. Using an approach directly analogous to the scattering theory-based derivation of the dielectric constant $\epsilon$ in $\vec{D}=\epsilon \vec{E}$ in electrodynamics of a material medium, one can derive the dispersion corrections to the neutron optical potential, which can be written in the form

\begin{equation}
    \label{eq:refractionindexSears}
     n^{\prime}=1-{{2\pi N b^{\prime}} \over {k^{2}}}[1+J^{\prime}+{{\pi N b^{\prime}} \over {k^{2}}}]
 \end{equation}

where the (dominant) real part of $J^{\prime}={{2\pi \rho b} \over {k}} \int \sin{(2kr)}[1-g(r)] dr$ for an isotropic medium, where $g(r)$ is the pair correlation function for the atoms in the material, $n^{\prime}$ is the real part of the neutron index of refraction with the multiple scattering correction, $b^{\prime}$ is the neutron scattering length with the multiple scattering correction, $\rho$ is the number density of atoms in the material, and $k$ is the incident neutron wave vector. 

The calculations of these additional terms in the theory of dispersion in neutron optics which leads to a modified expression for the index of refraction $n^{\prime}$ presented below were conducted years ago using many different techniques. The calculations performed in the 80s~\cite{Sea82, Now82b,Die81,Now82a} built upon much earlier work~\cite{Foldy1945, Lax1951, Ekstein1951, Lax1952, Ekstein1953, Lenk1975, Blaudeck1976} and were conducted within the framework of the traditional multiple scattering theory outlined above. A different calculational method~\cite{Warner1985} based on resummation of dominant subclasses of diagrams important for backscattering was seen to give equivalent results. Yet a third approach motivated by a desire to understand decoherence in neutron optics~\cite{Lanz1997} used a Lindblad operator treatment and also agrees with the results presented below.  All of these calculations restore consistency with the optical theorem and reduce in appropriate limits to the usual kinematic limit.

The parameters which control the size of these corrections are $kb$,  $kR$, and $b/d$ where $d$ is the separation between atoms in the medium. For slow neutrons of relevance to this work all of these parameters are typically of order $10^{-3}$ to $10^{-4}$. This is comparable to the accuracies of some of the scattering length measurements and so must be taken into account to make a valid comparison between the different measurements which have used different values of momentum $k$.

These small energy-dependent dispersive corrections~\cite{Sears85} could be measured using the technique described above. Using a hard core approximation for $g(r)$, one can estimate $dV/dE={2J_{0}^{2} \over (ka)^{4}}\sin{(ka)}[ka\cos{(ka)}-\sin{(ka)]}$ where $a$ is the hard core radius and $J_{0}= 2\pi\rho b a^{2}$. The typical size of this effect is about $10^{-3}V_{0}$ in the regime of interest. 

Note that one of the dispersion correction terms in the formula above  has a quadratic dependence on the neutron-nucleus scattering amplitude. This term can be thought of as a neutron optical equivalent of nonlinear optics in the interactions of electromagnetic waves in matter.

\section{Systematic Effects}

Our proposed experimental technique would search for the characteristic neutron energy dependence of the neutron-nucleus optical potential $dV/dE$ and exploit the neutron Fizeau effect and its characteristic dependence on the speed $w$ of the medium to eliminate the background phase shifts from constant neutron optical potential contributions. Other known physical effects might generate a nonzero $dV/dE$ which can in principle become a source of systematic error. 


Even in the absence of neutron-nucleus resonances, the neutron optical potential must possess an energy dependence due to the optical theorem. In the meV energy range emitted by slow neutron sources, the neutron-nucleus scattering amplitude has a real part that is typically much larger than the imaginary part. This follows in turn from probability conservation as embodied in the optical theorem of nonrelativistic scattering theory, $Im[f(\theta=0)]={k \sigma \over 4\pi}$ where $k$ is the neutron wave vector, $\sigma$ is the total cross section, and $f(\theta=0)$ is the forward scattering amplitude.  which implies that the scattering is dominated by the s-wave component of the partial wave expansion of the scattering amplitude. For this case, $f(\theta=0)=f$ is of order $R$ in magnitude and $\sigma=4\pi |f|^{2}$, so the optical theorem implies $Im[f]=k[Re(f)^{2}+Im(f)^{2}] \approx kR^{2}\approx 10^{-4} |f|$. Furthermore the neutron interferometric methods we propose to employ are only sensitive to the real part of $f$. We therefore expect these higher-order effects from potential scatting to be small in most cases. 



The contribution from the energy dependence of the tails of the n-A resonances that are far away from $E=0$  depends on the details of the resonance energies and widths of the particular nuclei. We can evaluate these resonance corrections using nuclear data. Extensive data on n-A resonances exists. Nuclei close to magic numbers nucleus possess especially low level densities near threshold in the meV regime, but nuclei with $A$ away from magic numbers can possess resonance tails from positive-energy resonances that will generate an energy dependent neutron optical potential. As an example for the specific case of natural isotopic abundance Sn: three of its isotopes have n-A resonances between $0-10$ eV~\cite{NNDC,Shibata2011}: $^{113}$Sn ($E=8.3$ eV, $\Gamma_{n}=4.5$ meV), $^{117}$Sn ($E=1.3$ eV, $\Gamma_{n}=0.00011$ meV, a p-wave resonance), and $^{119}$Sn ($E=6.2$ eV, $\Gamma_{n}=0.00148$ meV). Using the real part of the resonance formula above one sees that the contributions to $\Delta b/|b|$ from the residual neutron energy dependence of $b_\mathrm{res}$ for Sn over a $\delta E=10$ meV range starting at 0.5 meV is of order ${\Gamma_{n}\delta E} \over {E_\mathrm{res}^{2}}$, which does not exceed $10^{-6}$ for any of these resonance parameters.

One can wonder about whether or not there are any unexpected effects which might come from the presence of the neutron in a medium undergoing acceleration, since most of the practical methods to move the sample will also accelerate the matter. The passage of slow neutrons through an accelerating material medium should produce small energy changes in the neutron beam according to arguments using the equivalence principle~\cite{Kowalski1993, Kowalski1995, Littrell1996, Nosov1998} if the boundaries of the medium accelerate relative to the neutron beam. These effects are very small but have been resolved experimentally using measurements with ultracold neutrons for the case where the boundaries of the medium accelerate~\cite{Frank2006, Frank2008, Frank2011, Frank2012}. The fractional sizes of these neutron energy changes are much smaller on the slow and thermal neutron energy range. 

The neutron-nucleus weak interaction can generate a term in the neutron optical potential that depends on neutron energy. Usually the term in the neutron optical potential which is highlighted is the part coming from the parity-odd component of the forward scattering amplitude proportional to $\vec{s} \cdot \vec{p}$ where $\vec{s}$ is the neutron spin and $\vec{p}$ is the neutron momentum. In addition to this term, however, the weak interaction can also contribute a spin-independent term to the neutron-nucleus forward scattering amplitude that survives in the $q \to 0$ limit. In the absence of resonances, this term is typically smaller than that from the neutron-nucleus strong interaction by a factor of $10^{-7}$. Only in the case of the greatly-amplified parity violation effects which can occur on p-wave neutron resonances in heavy nuclei discussed above would be expected to generate observable effects.  

\section{Conclusion}

The insensitivity of slow neutrons to electromagnetic backgrounds and the ability to conduct sensitive interferometric measurements with slow neutrons in matter make them a good choice to search for delicate effects in the phase shift of neutrons in a medium. We have shown that it should be possible to exploit a special feature of the Fizeau effect for slow neutrons, namely the absence of a phase shift of the neutron amplitude as it passes through a medium moving parallel to its boundaries, to conduct a more sensitive search for various types of energy-dependent contributions to the neutron optical potential using recently-developed forms of neutron interferometry which can operate over a broad neutron energy range. We outlined scientific applications for this idea in the areas of neutron optics theory, nuclear data evaluation, neutron scattering data input for rare earth elements, and parity violation in low-energy p-wave neutron-nucleus resonances near threshold. Our proposed technique can be realized at existing neutron facilities and is especially convenient to implement at pulsed neutron sources. Possible systematic errors from other physical effects appear to be small and to possess calculable dependence on the velocity of the medium.

\section{Acknowledgements}
\label{sec:ack}

This research was funded by the Science Committee of the Ministry of Science and Higher Education of the Republic of Kazakhstan (Grant No. AP23484023). W. M. Snow acknowledges support from US National Science Foundation grant PHY-2209481 and from the Indiana University Center for Spacetime Symmetries. We acknowledge V. Gudkov for bringing our attention to issues in the analysis of negative energy resonance data.

\end{document}